\documentstyle[preprint,revtex]{aps}
\begin{document}
\draft
\preprint{\vtop{\baselineskip14pt\hbox{ETH-L}
                                 \hbox{(May 1996)}}}
\begin{title}
The Spinless Calogero-Sutherland model with twisted boundary condition\\ 
\end{title}
\author{D. F. Wang}
\begin{instit}
Institut de Physique Th\'eorique\\
Ecole Polytechnique F\'ed\'erale de Lausanne\\
PHB-Ecublens, CH-1015 Lausanne, Switzerland
\end{instit}
\begin{abstract}
In this work, the Calogero-Sutherland model with twisted boundary
condition is studied. The ground state wavefunctions, the ground 
state energies, the full energy spectrum are provided in details. 
\end{abstract}
\narrowtext

Exact solutions have provided us with important non-perturbative
insights in dealing with systems of strong correlations.
While there are very few exactly solvable systems available,
the ones that exist have yielded many interesting results.
Notable examples include electron systems with delta function
interactions\cite{yang1}, 
the Hubbard model\cite{wu}, the Kondo impurity spin 
system with linear conduction electron system\cite{andrei},
the Luttinger model\cite{mattis1} ,
and the Anderson model\cite{wiegmann}.
These models have all played important roles
in our understanding of physics in condensed matter theory.
Ever since Haldane and 
Shastry independently introduced the exactly solvable spin chain of $1/r^2$ 
exchange interaction\cite{haldane,shastry}, there has been considerable
activity in studying the variants of the Haldane-Shastry spin
chain doped with holes, {\it i.e.}\ the $t$-$J$ models of long range
hopping and exchange\cite{kura,kawa,wang1,wang2,wang3,hahaldane,ino,liu}.
It is interesting that the chiral Hubbard model\cite{gebhard}, which
at half-filling and in the limit of large but finite on-site energy 
reduces to the Haldane-Shastry spin chain, 
is also exactly 
solvable for any filling numbers and any on-site energy. 
In the following, we will study in details the Calogero-Sutherland model
with twisted boundary condition. The ground state wavefunctions, 
the ground state energies, and the full energy spectrum are provided.
Since one has to deal with the cases of bosons and fermions with twisted
boundary condition, the full discussion is divided into several sections as below.  

\section{The ground states}
\subsection{Spinless boson gas} 

We first consider the CS model of boson gas 
defined on a closed ring of length $L$. In the presence of a flux tube that threads
through the ring, the eigenenergy problem can be formulated as follows. 
Suppose that there are $N$ spinless bosons moving on the ring, $0\le x_i \le L; 
i=1,2,\cdots,N$.
Then the eigenvalue problem is
\begin{eqnarray}
H_{CS} &&\tilde 
\Psi(x_1\sigma_1, \cdots, x_i\sigma_i, \cdots, x_{N}\sigma_{N})\nonumber\\
&&=E
\tilde \Psi(x_1\sigma_1, \cdots, x_i\sigma_i, \cdots, x_{N}\sigma_{N}).
\end{eqnarray}
Here the Calogero-Sutherland Hamiltonian $H_{CS}$ takes the usual form
\begin{equation}
H_{CS} =- {1\over 2} \sum_{i=1}^N {\partial^2 \over \partial x_i^2}
+ \sum_{i < j} l(l+1)/[({L\over \pi})^2
\sin^2\left({\pi(x_i-x_j)\over L}\right)],
\end{equation}
where we assume $l>0$. 
The wavefunction obeys the twisted boundary condition
\begin{equation}
\tilde \Psi(x_1, \cdots, (x_i+L), \cdots, x_{N})
= e^{i\phi}
\tilde \Psi(x_1, \cdots, x_i, \cdots, x_{N}).
\label{eq:boundary}
\end{equation}
Obviously, the system is invariant under the translational operation
$\phi\rightarrow \phi+2\pi$.
Therefore we only need to consider the region $ -\pi \le \phi \le \pi$.
For the bosons,  
\begin{equation}
\tilde \Psi (x_1,\cdots,x_i,\cdots,x_j,\cdots)=\tilde \Psi (x_1,\cdots,x_j,\cdots,x_i,\cdots,x_N),  
\label{eq:symmetry1} 
\end{equation} 
i.e. the wavefunction $\tilde \Psi$ is symmetric under exchange of any two particles. 

Let us define the region $R$ as follows:
$\{R: 0\le x_i\le L; i=1,2,\cdots,N\}$. The sub-region of the full $R$
is denoted by $R_1: \{R_1: 0\le x_1\le x_2 \le \cdots \le x_N \le L\}$.  
The wavefunction inside the region $R_1$ is denoted 
by $\tilde \Psi_1(x_1,x_2,\cdots,x_N)$. The wavefunction  
in other sub-regions can be obtained by using the symmetry property of 
bosons Eq.(\ref{eq:symmetry1}). The twisted boundary condition Eq.(\ref{eq:boundary})
is translated to be
\begin{equation}
\tilde \Psi_1 (x_2,x_3,\cdots,x_N,L) = e^{i\phi} \tilde \Psi_1 (0,x_2,x_3,\cdots,x_N).
\end{equation} 
The ground state of the spinless bosons should take the following form
\begin{equation}
\tilde \Psi_1^g (x_1,x_2,\cdots,x_N) = e^{{i\phi\over L} \sum_{j=1}^N x_j}
\prod_{1\le i < j\le N} |\sin({\pi (x_i-x_j)\over L})|^{l+1}.
\label{equation:groundstate1} 
\end{equation}
One may check that the wavefunction satisfies the twisted boundary condition.
This wavefunction is also an eigenstate of the Hamiltonian. Since 
the wavefunction has no zeros in the region $R_1$, it is the ground state. 
The eigenenergy of the state can be found to be
\begin{equation}
E_g(\phi)= {1\over 2} N ({\phi\over L} )^2 + {1\over 6} (l+1)^2 \pi^2 N(N^2-1)/L^2. 
\end{equation}
Without the flux, the results reduces to those of Sutherland's\cite{sutherland2}. 
In the full space $R$, the ground state wavefunction $\tilde \Psi^g$ takes the 
simple form
\begin{equation}
\tilde \Psi^g(x_1,x_2,\cdots,x_N)=e^{{i\phi\over L} \sum_{j=1}^N x_j}
\prod_{1\le i < j\le N} |\sin({\pi (x_i-x_j)\over L})|^{l+1},
\end{equation}
which is symmetric under exchange of two particles, and which
satisfies the twisted boundary condition Eq.(\ref{eq:boundary}).
In presence of the flux, there is a persistent current in the ring. 
The persistent current is
$I(\phi)=-{\partial E_g(\phi) \over \partial \phi} = -{N\over L^2} \phi.$ 

\subsection{Spinless fermion gas (odd $N$)}

In this section, we discuss the spinless fermion model described 
by CS model in presence of magnetic flux tube. The eigenvalue
problem is formulated as follows:  
\begin{eqnarray}
H_{CS} &&\tilde
\Psi(x_1\sigma_1, \cdots, x_i\sigma_i, \cdots, x_{N}\sigma_{N})\nonumber\\
&&=E
\tilde \Psi(x_1\sigma_1, \cdots, x_i\sigma_i, \cdots, x_{N}\sigma_{N}), 
\end{eqnarray}
with the Calogero-Sutherland Hamiltonian $H_{CS}$ as before 
\begin{equation}
H_{CS} =- {1\over 2} \sum_{i=1}^N {\partial^2 \over \partial x_i^2}
+ \sum_{i < j} l(l+1)/[({L\over \pi})^2
\sin^2\left({\pi(x_i-x_j)\over L}\right)],
\end{equation}
where the coupling constant $l>0$.
The wavefunction  satisfies the twisted boundary condition
\begin{equation}
\tilde \Psi(x_1, \cdots, (x_i+L), \cdots, x_{N})
= e^{i\phi}
\tilde \Psi(x_1, \cdots, x_i, \cdots, x_{N}).
\label{eq:boundary2}
\end{equation}
In this case, since the system is made of spinless fermions, the wavefunction
is antisymmetric when exchanging two particles,
\begin{equation}
\tilde \Psi (x_1,\cdots,x_i,\cdots,x_j,\cdots)=(-1) \tilde \Psi (x_1,\cdots,x_j,\cdots,x_i,\cdots,x_N). 
\label{eq:symmetry2}
\end{equation}
As before, we define the full region $R$  to be $\{R: 0\le x_i\le L; i=1,2,\cdots,N\}$.
The sub-region $R_1$ is $\{R_1: 0\le x_1\le x_2 \le \cdots \le x_N \le L\}$.  
The twisted boundary condition of the wavefunction $\tilde \Psi (x_1,x_2,\cdots,x_N)$ defined in
$R$ is translated to a condition satisfied by the wavefunction $\tilde \Psi_1(x_1,x_2,\cdots,x_N)$
defined in the region $R_1$ as below:
\begin{equation}
\tilde \Psi_1 (x_2,x_3,\cdots,x_N,L) = 
(-1)^{(N-1)} e^{i\phi} \tilde \Psi_1 (0,x_2,x_3,\cdots,x_N).
\label{eq:boundary10}
\end{equation}
Given $\tilde \Psi_1$ inside $R_1$, the wavefunctions in other sub-regions of $R$ can be obtained
using the anti-symmetry of the fermionic statistics. 
Since $N$ is odd, the prefactor $(-1)^{(N-1)}$ disappears. We propose the following
wavefunction as the ground state: inside $R_1$, the ground state takes Jastrow form
\begin{equation}
\tilde \Psi_1^g (x_1,x_2,\cdots,x_N) = e^{{i\phi\over L} \sum_{j=1}^N x_j}
\prod_{1\le i < j\le N} |\sin({\pi (x_i-x_j)\over L})|^{l+1}.
\label{equation:groundstate2}
\end{equation}
One may compute the corresponding eigenvalue of this wavefunction. It is found
that 
\begin{equation}
E_g(\phi)= {1\over 2} N ({\phi\over L} )^2 + {1\over 6} (l+1)^2 \pi^2 N(N^2-1)/L^2.
\end{equation}
In presence of the magnetic flux, there is a persistent current in the system. 
The persistent current is
$I(\phi)=-{\partial E_g(\phi) \over \partial \phi} = -{N\over L^2} \phi$.
Without the flux, our ground state wavefunction reduces to that of Sutherland's. 
In the full region $R$, the ground state wavefunction $\tilde \Psi^g$ can be written
in a compact way
\begin{equation}
\tilde \Psi^g (x_1,x_2,\cdots,x_N) = e^{{i\phi\over L} \sum_{j=1}^N x_j}
\prod_{1\le i < j\le N} |\sin({\pi (x_i-x_j)\over L})|^l \times 
\sin({\pi (x_i-x_j)\over L}),
\end{equation} 
which is anti-symmetric when exchanging two fermions, and which also satisfies 
the twisted boundary condition Eq.(\ref{eq:boundary10}). 

\subsection{Spinless fermion Gas (even $N$)}

The eigenvalue problem is formulated as before. However, the boundary condition  
Eq.(\ref{eq:boundary10}) should be taken great care of. First, let us consider 
the situation without flux, $\phi=0$, and we impose periodic boundary condition
on the wavefunction 
\begin{equation}
\tilde \Psi(x_1, \cdots, (x_i+L), \cdots, x_{N})
=\tilde \Psi(x_1, \cdots, x_i, \cdots, x_{N}).
\end{equation}
This PBC is translated to be a boundary condition for $\tilde \Psi_1$ as below:
\begin{equation}
\tilde \Psi_1 (x_2,x_3,\cdots,x_N,L) = (-1)\tilde \Psi_1 (0,x_2,x_3,\cdots,x_N).
\end{equation}
With this in mind, the ground state for the system inside the region $R_1$ should
take the following form:
\begin{equation}
\tilde \Psi_1^g (x_1, x_2, \cdots,x_N) =e^{{\pm i\pi\over L}\sum_{j=1}^N x_j} 
\times \prod_{1\le i < j\le N} |\sin({\pi (x_i-x_j)\over L})|^{l+1}.
\end{equation} 
We can compute the eigen-energy of this wavefunction. The ground state energy
is found to be 
\begin{equation}
E_g={1\over 2} N ({\pi\over L})^2 +{1\over 6} (l+1)^2 \pi^2 N(N^2-1)/L^2.
\end{equation} 
This energy is different from the bosonic case, as well as different from 
the fermionic case when the total number of particles is odd.  
 
Now, let us consider the situation when there is nozero flux.
Consider the case where $0\le \phi \le \pi$.  
The twisted boundary condition Eq.(\ref{eq:boundary10}) for the wavefunction
$\tilde \Psi_1$ is satisfied by the following Jastrow product
\begin{equation}
\tilde \Psi_1^g (x_1, x_2, \cdots,x_N) =e^{{i(\phi-\pi)\over L}\sum_{j=1}^N x_j}
\times \prod_{1\le i < j\le N} |\sin({\pi (x_i-x_j)\over L})|^{l+1}.
\end{equation} 
This wavefunction is the ground state of the system, with the ground state energy
given by
\begin{equation}
E_g (\phi)={1\over 2} N ({\pi-\phi\over L})^2 +{1\over 6} (l+1)^2 \pi^2 N(N^2-1)/L^2.
\end{equation}
The persistent current of the system is found to be 
$I(\phi)=-\partial E(\phi)\partial \phi = -{N\over L^2} (\phi-\pi) \ge 0$. 
If the system has a flux $-\pi \le \phi \le 0$, 
the ground state wavefunction is found to be 
\begin{equation}
\tilde \Psi_1^g (x_1, x_2, \cdots,x_N) =e^{{i(\phi+\pi)\over L}\sum_{j=1}^N x_j}
\times \prod_{1\le i < j\le N} |\sin({\pi (x_i-x_j)\over L})|^{l+1}.
\end{equation}
The corresponding eigen-energy is given by
\begin{equation}
E_g (\phi)={1\over 2} N ({\pi+\phi\over L})^2 +{1\over 6} (l+1)^2 \pi^2 N(N^2-1)/L^2.
\end{equation}
The persistent current is $I(\phi)=-{N\over L^2} (\phi+\pi) \le 0$. 

In the full space $R$ as defined before, the ground state wavefunction
$\tilde \Psi^g$ can be written in a compact way. 
For $0\le \phi \le \pi$, inside the full region $R$, the ground state 
wavefunction is 
\begin{equation}
\tilde \Psi^g (x_1, x_2, \cdots,x_N) =e^{{i(\phi-\pi)\over L}\sum_{j=1}^N x_j}
\times \prod_{1\le i < j\le N} |\sin({\pi (x_i-x_j)\over L})|^l
\times \sin({\pi (x_i-x_j)\over L}).
\end{equation}
While for $-\pi\le \phi \le 0$, inside $R$, one has 
\begin{equation}
\tilde \Psi^g (x_1, x_2, \cdots,x_N) =e^{{i(\phi+\pi)\over L}\sum_{j=1}^N x_j}
\times \prod_{1\le i < j\le N} |\sin({\pi (x_i-x_j)\over L})|^l
\times \sin({\pi (x_i-x_j)\over L}).
\end{equation} 
In next section, we will provide the full energy spectrum for the CS model
under twisted boundary condition, following a similar approach
of Sutherland's for zero flux case\cite{sutherland2}. 

\section{Excitation spectrum} 

\subsection{Spinless boson gas}

Following the idea of Sutherland\cite{sutherland2}, one can write the wavefunction
as a product of the Jastrow part and the part of plane waves. 
Keeping the plane waves due to the twisted boundary condition
we can find the energy spectrum of 
the spinless boson gas given by
\begin{equation}
E={\pi^2\over 6} (l+1)^2 N(N^2-1)/L^2 + {1\over 2} (2\pi/L)^2 \epsilon 
\end{equation}
where the function $\epsilon$ is given by
$\epsilon= \sum_{j=1}^N (n_j+ {\phi\over 2\pi})^2 + (l +1) \sum_{i>j}
[n_i-n_j].$ The quantum numbers $n_j$ are non-negative integers, which satisfy  
condition $n_{j+1} \ge n_j$.  
The quantum numbers do not have to be distinct from each other. The ground state is 
obtained when all $n_j=0$. 

\subsection{Spinless fermions (odd $N$)}

For the spinless fermion gas (odd $N$), one can also find the 
excitation spectrum of the system under the twisted boundary condition.
The full energy spectrum takes the form 
\begin{equation}
E={\pi^2\over 6} (l+1)^2 N(N^2-1)/L^2 + {1\over 2} (2\pi/L)^2 \epsilon
\end{equation}
where the function $\epsilon$ is given by
$\epsilon= \sum_{j=1}^N (n_j+ {\phi\over 2\pi})^2 + (l+1) \sum_{i>j}
[n_i-n_j].$ The quantum numbers $n_j$ are non-negative integers, and one has  
the condition $n_{j+1} \ge n_j$.   
The ground state is given when all $n_i=0$. 

\subsection{Spinless fermions (even $N$)}

Finally, for the spinless fermion gas of even $N$, we also find the 
full energy spectrum taking the following form for $ 0 \le \phi \le \pi$. 
\begin{equation}
E={\pi^2\over 6} (l+1)^2 N(N^2-1)/L^2 + {1\over 2} (2\pi/L)^2 \epsilon
\end{equation}
where the function $\epsilon$ is given by
$\epsilon= \sum_{j=1}^N (n_j+ {\phi-\pi\over 2\pi})^2 + (l+1) \sum_{i>j}
[n_i-n_j].$ The quantum numbers $n_j$ are non-negative integers that have  
the condition $n_{j+1} \ge n_j$.   
The ground state corresponds to all $n_i=0$. 

When the flux $ -\pi \le \phi \le 0$, the energy excitation is given by
\begin{equation}
E={\pi^2\over 6} (l+1)^2 N(N^2-1)/L^2 + {1\over 2} (2\pi/L)^2 \epsilon
\end{equation}
where the function $\epsilon$ is given by
$\epsilon= \sum_{j=1}^N (n_j+ {\phi+\pi\over 2\pi})^2 + (l+1) \sum_{i>j}
[n_i-n_j].$ The quantum numbers $n_j$ are non-negative integers, satisfying
the condition $n_{j+1} \ge n_j$.   
The ground state is reached when all $n_i=0$. 
 
\section{Summary}

In summary, we have discussed how the boundary condition affects the 
spinless CS model of long range interaction. The ground state wavefunctions,
the ground state energies, the full energy spectrum are provided for
both the fermionic gas and the bosonic gas. The exact solutions indicate 
that the parity effect
for the persistent currents still hold for the fermionic gas, in spite 
of the electron-electron correlation.  

I wish to thank C. Gruber, J. T. Liu, C. A. Stafford, H. Kunz and X. Q. Wang
for conversations. 
This work was supported 
by the Swiss National Science Foundation.

\end{document}